# The importance and need for system monitoring and analysis in HPC operations and research

Florina M. Ciorba

Department of Mathematics and Computer Science
University of Basel, Switzerland

In this work, system monitoring and analysis are discussed in terms of their significance and benefits for operations and research in the field of high performance computing (HPC). HPC systems deliver unique insights to computational scientists from different disciplines. It is argued that research in HPC is also computational in nature, given the massive amounts of monitoring data collected at various levels of an HPC system. The vision of a comprehensive system model developed based on holistic monitoring and analysis is also presented. The goal and expected outcome of such a model is an improved understanding of the intricate interactions between today's software and hardware, and their diverse usage patterns. The associated modeling, monitoring, and analysis challenges are reviewed and discussed. The envisioned comprehensive system model will provide the ability to design future systems that are better understood before use, easier to maintain and monitor, more efficient, more reliable, and, therefore, more productive. The paper is concluded with a number of recommendations towards realizing the envisioned system model.

## 1 Introduction

Each of the four pillars (experiments, theory, simulation, and data) of the scientific method produce and consume large amounts of data. Breakthrough science will occur at the interface between empirical, analytical, computational and data-based observation. Parallel computing systems are the workhorse of the third pillar: simulation science. These systems are highly complex ecosystems, with multiple layers, ranging from the hardware to the application layer (Fig. 1). Monitoring solutions exist at every single layer. Access to the monitoring data varies between the different communities, which also have different interests in the data. For instance, computational scientists in general have access to the monitoring data from the application layer and in certain cases also from the application environment layer. Their interests may include the running time of their applications but also understanding the application performance by profiling or tracing. Computer scientists may have access to monitoring data from application environment and cluster software layers. They are typically interested in performance data from both software and hardware components and subsystems. System administrators typically have access to data monitored at the hardware, system software, and cluster software layers. The interests regard more the operational aspects of the system, including availability, fair usage, etc. Accessing and sharing of data based on the actual community interests (as depicted by the arrows in Fig. 1) may require bilateral agreements between the communities or public release of data.

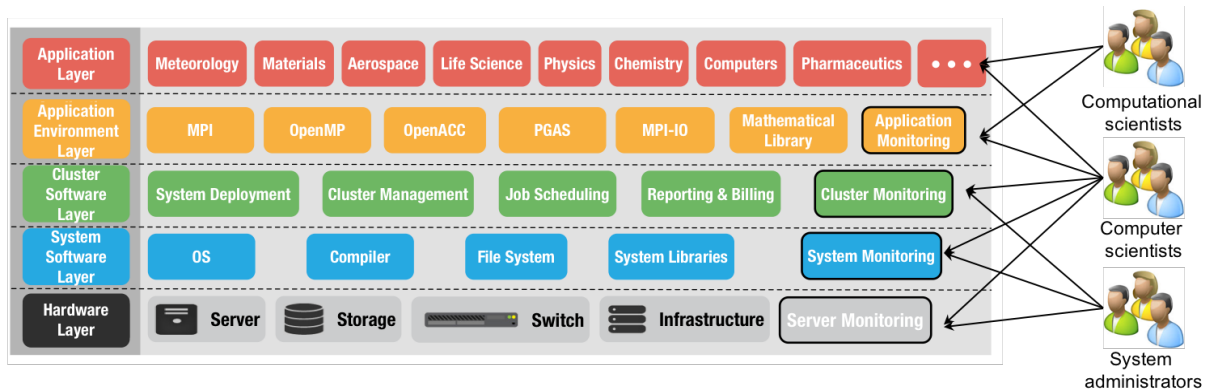

Figure 1: The HPC ecosystem and the monitoring interests of the different communities.

Understanding the interaction between the software running on parallel computers and the underlying hardware is a notorious challenge, especially given the fact that the hardware models and designs change very often. This is due to the rapid advances in computing technologies. Studying the interaction between hardware and software in this context will require integration of monitoring solutions from the different system layers. System monitoring relates to data science (the fourth pillar) in the sense that monitoring applications and systems generates a large amount of data from various sources, at various rates, and under various formats. This brings about all the data-related challenges also faced in other domains that require a data science approach. Thus, scientific research in high performance computing that involves very large amounts of monitoring data becomes *computational computer science.* Conducting research in computational computer science is challenging due to the fact that the "physical world" to be modeled is changing with every new software and hardware release. Specifically, modeling system behavior is a significant challenge for the following reasons: (1) The hardware architecture is heterogeneous. (2) The system stack is also heterogeneous. (3) There exist multiple types of execution environment.

This paper describes the importance and need of holistic system monitoring in view of developing a comprehensive model of the system behavior, wherein "system" denotes both hardware and software components. It is argued that such monitoring and modeling will have positive impacts and ramifications in HPC operations and research, not only for existing systems, but most importantly for the design, procurement, operation, productivity, and maintenance of future HPC systems. A review of existing monitoring and analysis solutions of the different system layers is also presented. The envisioned system model requires a comprehensive view of the applications and the system that captures and integrates activities from all system layers (Fig. 1).

## 2 Current Monitoring and Analysis Solutions

The process of monitoring and analysis consists of three stages: data collection, data processing, and data visualization and analysis. Monitoring and analysis is employed at various layers of the HPC ecosystem, as illustrated in Fig. 1. Below is a brief review of monitoring solutions for each layer, in a top-down manner, followed by pointers to system-wide monitoring and analysis efforts. The purpose of this review is rather illustrative and not meant to be comprehensive.

### 2.1 Single-Layer Monitoring and Analysis

**Application Layer** The monitoring and analysis solutions described herein pertain both to the *application* and *application environment* layers (Fig. 1). The Virtual Institute for High Productivity Supercomputing (VI-HPS) also provides a great overview and guide in selecting application level productivity tools[1].

*A. Automatic profile and trace analysis*: The Periscope tuning framework[2] enables automatic analysis and tuning of parallel applications. TAU[3] is a set of tuning and analysis utilities that offer an integrated parallel performance system. Score-P[4] is a community-developed instrumentation and measurement infrastructure, targeting both performance and energy. Extra-P[5]: automatic performance-modeling tool that supports the user in the identification of scalability bugs. mpiP[6] is an MPI profiling tool, while mpiPview[7] is an MPI performance profile analysis and viewer. Open|SpeedShop[8] is an integrated parallel performance analysis environment. IPM[9] is an integrated performance monitoring profiling tool.

*B. Visualization and analysis*: Vampir[10] is an interactive graphical trace visualization and analysis tool supporting identification of performance hotspots. Paraver/Dimemas/Extrae[11] form a suite of tools allowing the profiling, tracing and sampling of an application, supported by graphical trace visualization and analysis. Scalasca[12] a tool for large-scale parallel program performance analysis supporting identification of performance bottlenecks. Allinea PR[13] provides performance reports that show the behavior of parallel HPC application runs.

*C. Optimization*: Rubik[14] is a tool that generates topology-aware mapping of processes, optimized for IBM Blue Gene machines. MAQAO[15] is a solution for analyzing and optimizing binary codes on single nodes, supporting Intel64 and Xeon Phi architectures.

*D. Debugging, error and anomaly detection*: MUST[16] is a runtime error detection tool for MPI applications Archer[17] is a low overhead data race detector for OpenMP programs. STAT[18] is a tool for identifying errors in code running at full system scale. Allinea Forge tools[19] includes DDT and MAP, and provides a single interface for debugging, profiling, editing and building code on local and remote systems. RogueWave TotalView[20] is a parallel debugging tool that allows fault isolation, memory optimization, and dynamic visualization for high-scale HPC applications.

**Cluster Layer** The solutions employed at cluster level are either used for cluster management or job scheduling and reporting.

---

[1] http://www.vi-hps.org/tools/
[2] http://periscope.in.tum.de
[3] http://www.cs.uoregon.edu/research/tau/
[4] http://www.score-p.org
[5] http://www.scalasca.org/software/extra-p
[6] https://computing.llnl.gov/code/mpip-llnl.html
[7] https://computation.llnl.gov/casc/tool_gear/mpipview.html
[8] https://openspeedshop.org/
[9] http://ipm-hpc.sourceforge.net
[10] https://www.vampir.eu
[11] http://www.bsc.es/computer-sciences/performance-tools
[12] http://www.scalasca.org
[13] https://www.allinea.com/products/allinea-performance-reports
[14] https://github.com/LLNL/rubik
[15] http://www.maqao.org
[16] http://www.itc.rwth-aachen.de/MUST
[17] https://github.com/PRUNER/archer
[18] http://www.paradyn.org/STAT/STAT.html
[19] https://www.allinea.com/products/develop-allinea-forge
[20] http://www.roguewave.com/products/totalview.aspx

*A. Cluster management*: Nagios[21] represents the industry standard in IT infrastructure monitoring, including critical components, log server, and network. Ganglia[22] is a scalable and distributed monitoring system used in approximately 90% of HPC shops. It represents data in XML, uses XDR for compact and portable data transfer, and stores and visualizes data using the RRDtool. Supermon[23] is a high-speed cluster monitoring system, that emphasizes low perturbation and high sampling rates. Zenoss[24] provides event-based IT monitoring and analytics for cloud, virtual and physical IT environments. Cerebro[25] is a collection of cluster monitoring tools and libraries, not yet scaling to a full host level. Splunk[26] is a collection of software for searching, monitoring, and analyzing machine-generated big data, via a web-style interface and team dashboards. RUR[27] (resource utilization reporting) is an administrator tool for gathering statistics on how system resources are being used by applications, developed by Cray for its HPC systems. RUR runs before or after applications collecting system statistics.

*B. Job scheduling and reporting*: SysMon[28] is a Windows-based system service that logs system activity in XML format; it is also a batch system monitor plugin for Eclipse PTP. Other batch scheduling systems (e.g., SLURM, UNIVA Grid Engine, PBS/Pro, LSF, Moab/Torque) provide statistics from batch jobs by running daemons on the compute nodes during application execution.

**System Layer** Current solutions for system monitoring and analysis mainly target the operating system and storage/file system behavior.

*A. Operating system*: Collectd[29] is a system statistics collection daemon, which collects performance metrics periodically and can interface with Nagios (see above) and LMT (see below). Perf[30] is a lightweight profiling command line tool for Linux 2.6+ systems (included in the kernel tools); it abstracts CPU hardware differences. Collectl[31] is a comprehensive OS-level monitoring tool which gathers information about Linux system resources: CPU, memory, network, inodes, processes, nfs, tcp, sockets, and others. ProcMon[32] is an advanced monitoring tool for Windows (part of Sysinternals), which shows real-time file system, Registry and process/thread activity.

*B. Storage/File system*: LMT (Lustre Monitoring Tool, http://wiki.lustre.org/Lustre_Monitoring_Tool) is an open-source tool for capturing and displaying of Lustre file system activity. It monitors Lustre file system servers, collects data using the Cerebro monitoring system in a MySQL database. Lltop (https://github.com/jhammond/lltop is a command line utility that monitors file system load (Lustre I/O statistics) and is integrated with job assignment data from cluster batch schedulers. GPFS Monitor Suite (https://sourceforge.net/projects/gpfsmonitorsuite/) uses Perl scripts and Ganglia to monitor IBM's general parallel file system (GPFS) multi-cluster environments, and can interface with Nagios.

**Server Layer** The solutions employed at the server level target the different node-level (and not only) subsystems: CPU, memory, network, I/O, and storage, respectively.

*A. Node/CPU*: PAPI[33] (performance API) is an interface to hardware performance counters found in most microprocessors. Syslog[34] is a Linux system logging utility, which has become standard for message logging, applied to various facilities: kernel, user, mail, printer, network, and others.

---

[21] https://www.nagios.org
[22] http://ganglia.info
[23] http://supermon.sourceforge.net
[24] https://www.zenoss.com/
[25] https://github.com/chaos/cerebro
[26] https://www.splunk.com/
[27] http://pubs.cray.com/#/00241000-FA/FA00240785
[28] http://ss64.com/nt/sysmon.html
[29] https://collectd.org
[30] https://perf.wiki.kernel.org/
[31] http://collectl.sourceforge.net
[32] https://technet.microsoft.com/en-us/sysinternals/bb896645.aspx
[33] http://icl.cs.utk.edu/papi/
[34] https://linux.die.net/man/8/syslogd

*B. Memory*: Callgrind[35] is an open-source profiler using execution-driven cache simulation. It is accompanied by KCachegrind, a visualization GUI that provides various views of performance data (annotated call graphs and tree maps, annotated source and machine code). Memchecker[36] a framework integrated within Open MPI that finds hard-to-catch memory errors in parallel applications.

*C. Network*: Monitoring of fabric (e.g., InfiniBand, OmniPath) counters is performed with operating system-level tools such as collectl (see above). Pandora FMS, OpenNMS, NetXMS, and Zabbix are open source network management systems which also perform network performance monitoring. They have recently been compared in a recent article published by Network World[37].

*D. I/O*: SIONlib[38] is a scalable I/O library for parallel access to task-local files that uses shared files for task-local data, written through collective I/O operations. Darshan (http://www.mcs.anl.gov/research/projects/darshan/) is a characterization tool of the I/O behavior of scalable HPC applications. IOTA[39] (I/O tracing and allocation library) is a lightweight scalable tracing tool for diagnosing poorly performing I/O operations to parallel file systems, especially Lustre.

*E. Storage*: DirectMon[40] is a configuration and monitoring solution for DDN storage devices. Iostat[41] is part of the sysstat family of Linux tools, which reports statistics for CPU, input and output storage devices, partitions, and network file systems.

**Remarks** Most of the solutions described above are released under open-source licenses while certain are licensed commercially (e.g., tools from Allinea, Splunk, Zenoss). The majority of the tools are also interoperable (e.g., those included in the VI-HPS Tools Guide[42]). One can observe that more and more tools are being developed and released per layer over time. Truly needed, however, are efforts that integrate these or other tools ideally across all four layers (with respect to Fig. 1), such that a *holistic view* of the behavior of the applications, the software they employ, as well as the hardware on which they run can become feasible.

## 2.2 System-Wide Monitoring and Analysis

The following efforts employ, to different extents, solutions for system-wide or, in some cases, integrated monitoring and analysis of HPC applications and systems.

Elastic Stack[43] consists of Elasticsearch, which along with Logstash and Kibana, provides a powerful platform for indexing, searching and analyzing the data stored in logs. It also includes Beats, an open source platform for building lightweight data shippers for log files, infrastructure metrics, network packets, and more.

Graylog[44] is a powerful open source log management and analysis platform, that has many use cases, from monitoring SSH logins and unusual activity to debugging applications. It is based on Elasticsearch, Java, MongoDB, and Scala.

---

[35] http://kcachegrind.sourceforge.net/html/Home.html
[36] http://www.open-mpi.org/
[37] http://www.networkworld.com/article/3142061/open-source-tools/pandora-fms-wins-open-source-management-shootout.html
[38] http://www.fz-juelich.de/SIONlib
[39] https://bitbucket.org/mhowison/iota
[40] www.ddn.com/products/storage-management-directmon/
[41] https://linux.die.net/man/1/iostat
[42] http://www.vi-hps.org/upload/material/general/ToolsGuide.pdf
[43] https://www.elastic.co
[44] https://www.graylog.org

Perfsuite[45] is an accessible, open source performance analysis environment for Linux. It consists of a collection of tools, utilities, and libraries for software performance analysis that are easy to use, comprehensible, interoperable, and simple. It serves as starting point for more detailed analyses and for selection of other more specialized tools/techniques.

The SUPReMM[46] project targeted integrating monitoring and modeling of HPC systems usage and performance of resources of XSEDE cluster, and was jointly carried out between the University at Buffalo and Texas Advanced Computing Center (TACC) from 2012-2015. One outcome of this project is TACC stats.

TACC Stats[47] is a package providing tools for automated system-wide resource-usage monitoring and analysis of HPC systems at multiple levels of resolution.

Decision HPC[48] is an analytics and monitoring tool developed by X-ISS, hooks into existing HPC schedulers (e.g., Torque, PBS Pro, LSF, CJM, and Univa Grid Engine) and can interface with common (e.g., Ganglia) or custom cluster monitoring tools. It displays a live dashboard for real-time monitoring in a single pane of glass view.

OVIS[49] is an open source modular software system for data collection, transport, storage, analysis, visualization, and response in HPC systems under active development at Sandia National Laboratories and Open Grid Computing[50]. It provides real-time, intelligent monitoring of the performance, health and efficiency of large scale computing systems.

"Holistic, Measurement-Driven Resilience: Combining Operational Fault and Failure Measurements and Fault Injection for Quantifying Fault Detection and Impact"[51] (HDMR) is a collaboration between the University of Illinois at Urbana-Champaign; Sandia, Los Alamos, and Lawrence Berkeley National Laboratories; and Cray Inc. The goal is to determine fault-error-failure paths in extreme scale systems of today, to help improve those and future systems.

The NSF project "Computer System Failure Data Repository to Enable Data-Driven Dependability"[52] (CSFDR) (2015-2018) will address the need for an open failure data repository for large and diverse computing infrastructures that will provide enough information about the infrastructures and the applications that run on them.

**Remarks** All these efforts address the challenge of system-wide monitoring and analysis in different manners. System-wide monitoring is not necessarily equivalent to *comprehensive* or *holistic* system monitoring. The failure data repository (CSFDR) and resiliency modeling (HMDR) projects take a comprehensive and holistic, respectively, approach to fault monitoring and modeling. However, modeling faulty behavior alone may leave out significant aspects such as logical reconfigurability, heterogeneity, performance variability, utilization of resources, and others. Therefore, all above efforts are a strong source of inspiration and motivation for considering all aspects of applications and system behavior holistically.

## 3 Modeling System Behavior based on Holistic System Monitoring

The author's vision is that system behavior can be modeled comprehensively based on insights and information obtained by holistic system monitoring. In this context, *system* is understood as the entirety of all its software (the four upper layers in Fig. 1) and hardware (the bottom layer in Fig. 1) components. Studying system behavior based on single-layer monitoring data limits the insight to the extent corresponding to that particular layer. A model of *system behavior* captures the intricate relation between the behavior of the components in each layer and across layers (Fig. 2). This intricate relation can only be observed and understood, by a "diagnosis team", through *holistic system monitoring* (Fig. 3). Such monitoring will offer a holistic view of the system and facilitate the comprehensive understanding of the system behavior.

---

[45] http://perfsuite.ncsa.illinois.edu
[46] https://github.com/ubccr/supremm
[47] https://github.com/TACC/tacc_stats
[48] https://www.hpcwire.com/2014/09/18/modernizing-hpc-cluster-monitoring/
[49] https://ovis.ca.sandia.gov/mediawiki/index.php/Main_Page
[50] https://www.opengridcomputing.com
[51] http://portal.nersc.gov/project/m888/resilience/
[52] https://www.nsf.gov/awardsearch/showAward?AWD_ID=1513197

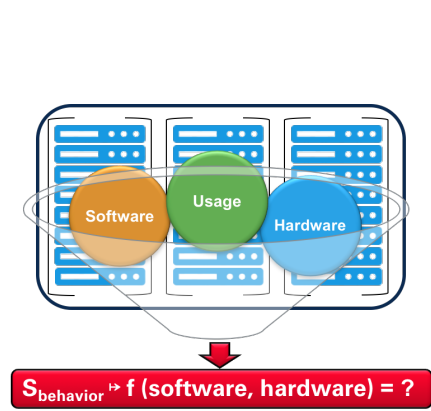
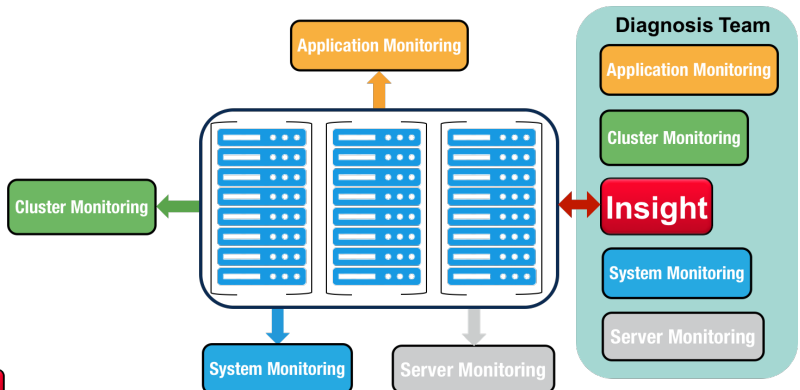

Figure 2: Vision of a system model.

Figure 3: Holistic system monitoring.

The envisioned system model will allow the systematic investigation of the interaction between hardware, software and operating conditions in high performance computers. Moreover, it will provide the ability to study the behavior and efficiency of high performance computers on application, cost, performance, and operational levels.

### 3.1 Challenges

**Modeling** Capturing the complex behavior of computer systems in a *model* is highly challenging because the "physical world" to be modeled is always changing, with every new software and hardware release. Since the interaction between newly released software and/or hardware components cannot be foreseen *a priori*, this interaction needs to be studied by observation *in situ*. In this context, the integration of monitoring solutions from the different system layers is the only path forward. Verification and validation of computer system models is a challenging and rewarding task. The same holds for the verification and validation of the envisioned system model, and will require experiments, simulation, and analytical work.

**Monitoring** Applications and systems *monitoring* generates large amounts of data from various sources (i.e., the system layers of Fig. 1), at various rates, and under various formats.

The data generated by different layers is "owned" by the different communities. For instance, application performance data is "owned" by the computational scientists, while system performance data is "owned" by the system administrators. Computer scientists may "own" data from the application environment layer. However, accessing data from higher or lower layers is not a trivial pursuit. In most cases, such data can be accessed either based on bilateral agreements between the computer scientists and computational scientists/system administrators, or if the data is publicly available.

Holistic monitoring and analysis requires a unified real-time view of all cluster(s) activities, presented in a comprehensive way. This implies that monitoring data needs to be gathered from application, cluster, system, and server levels, and integrated accordingly. In addition to the data ownership challenge mentioned earlier, data processing challenges include filtering, de-noising, curating, and de-identifying all the data, in view of extracting system behavior.

In terms of infrastructure, larger computing systems require correspondingly larger monitoring infrastructure(s). Consequently, the following questions arise naturally: (1) How can monitoring infrastructures be budgeted and funded? (2) How and where is the monitoring infrastructure hosted? (3) How does the quality of the monitoring hardware influence the quality of the monitoring information? (3) Does the monitoring infrastructure need any form of monitoring? (4) How can the monitoring data be collected on shared resources (e.g., network, storage, etc.) without requiring additional investments? (5) Can comprehensive monitoring be realized in a light-weight manner, to ensure minimal perturbation in system behavior?

**Analysis** The goal of monitoring applications and systems is to gain insight into their behavior by *analyzing* the empirically collected data, ultimately resulting in an analytical behavioral model. The data analysis process faces the following challenges: (1) De-identification and anonymization of sensitive data (e.g., user identifiers and activities, malfunctions, and security vulnerabilities). (2) Extracting value from monitoring data, such as job and project level statistics. This challenge is closely coupled with the choice of relevant metrics, which can be different for across systems. (3) Support of various types of analyses (e.g., qualitative, quantitative). (4) Correlation of data sources, in the form of e.g., integrated databases, multiple databases due to multiple data types. (5) Storing and near-real time indexing of monitoring metrics and logs storing. (6) Concerns about bandwidth consumption on shared resources and denial of service attacks.

**Other Challenges** There are several other challenges on the path to modeling system behavior through holistic monitoring and analysis. (1) Finding the best monitoring and analysis solution for a given HPC system, given that not all comprehensive monitoring solutions will work in practice for all HPC systems. (2) Diagnosing system behavior (including performance) without the right data available is difficult. (3) Data collection and processing noise needs to be minimum (e.g., system logs are quite "noisy" as they collect most of the activities).

## 3.2 Importance and Value

**Importance** Comprehensive system monitoring is important as it provides the basis for maintaining good management and facilitating improvements of HPC systems operations. Understanding the current usage and behavior of HPC systems is the prerequisite for designing and purchasing future systems.

**Value and Benefits** Comprehensive system monitoring can be analogized to personalized system health and has a high value for the HPC stakeholders. More productive HPC operations and research at decreased costs is for the benefit of all scientists, system owners, administrators, and sponsors.

## 3.3 Expected Outcome and Impact

**Outcome** The expected outcome of the envisioned system mode based on comprehensive monitoring and analysis is the ability to (1) Give a personalized system diagnosis. (2) Identify application and system performance variation. (3) Detect contention for shared resources and assess impacts. (4) Discover misbehaving users and system components. (5) Reduce operational cost. (6) Improve system reliability. Ideally, these analyses should be automated. The system model will offer an improved understanding of the intricate interactions between today's software and hardware, and their diverse usage patterns.

**Impact** The impact of holistic system monitoring and analysis includes: a broader application and project performance assessment, historical trends of applications and systems, estimates of job and operational costs, as well as a positive impact on more targeted future HPC system procurements.

The impact of a comprehensive system model is providing the ability to *design future systems* that are: (1) Better understood before they are deployed and released for production use. (2) Easier to maintain and monitor, right from the start, decreasing the operational and maintenance costs. (3) Larger, more efficient (in terms of performance and energy), more reliable, and, therefore, more productive.

## 4 Summary and Conclusions

**Summary** System monitoring and analysis are important topics of high relevance to a wide range of stakeholders, ranging from system administrators to sponsors, and including computer and computational scientists, that share different interests. The importance of these topics is also supported by the increasing number of workshops, bird-of-a-feather session, panels and discussion events organized at different venues, as well as by the large ongoing and several new funding grants. Remarkable monitoring and analysis efforts exist for different system layers, as well as several system-wide efforts. In this paper, a vision for a system model describing the behavior and interaction between the software and hardware components is presented. The model is based on insights and information obtained by holistic system monitoring. Such monitoring offers a holistic view of the system to a "diagnosis team" which analyzes the information and captures the system behavior in a comprehensive model. The envisioned system model will allow systematic investigation of the interaction between hardware, software and operating conditions in high performance computers. Moreover, it will provide the ability to study the behavior and efficiency of high performance computers on application, cost, performance, and operational levels. A number of challenges have been identified regarding holistic system monitoring as well as comprehensive system modeling. The importance, value, expected outcome, and impact of the envisioned model have also been outlined.

**Recommendations** To conclude the paper, a set of recommendations for going forward are enumerated below. (1) Transform all challenges described herein as opportunities to gain better systems insights that will lead to better systems designs and happier stakeholders. (2) Include the design of the monitoring infrastructure and associated costs as part of the system design, procurement, and maintenance process. Monitoring is often excluded from this process, and seldom designated for performance. (3) Monitor both software and hardware. (4) Decide what aspects are important to record and analyze. (5) Perform a holistic, versus a disjoint, analysis of the monitoring data originating both in software and hardware. (6) Understand the behavior of current (smallest-to-largest) systems before building and using the next (largest) system.

## Acknowledgements

The author acknowledges Prof. Wolfgang E. Nagel, Director of the Center for Information Services and High Performance Computing at TU Dresden, for his support in establishing a system monitoring infrastructure on Taurus[53]. Thanks also go to Dr. Sadaf Alam, Associate Director at the Swiss National Supercomputing Centre, for establishing a system monitoring infrastructure to collect monitoring data on CSCS machines[54]. Acknowledgements also go to Dr. Ann Gentile and Dr. Jim Brandt for building a community site for monitoring large-scale HPC systems collaborations[55] and exchange of ideas. The author further acknowledges the HPC Team, Rechenzentrum, Universität Freiburg, lead by Prof. Gerhard Schneider for welcoming the idea of monitoring the bwForCluster NEMO[56] in view of collecting failure information and optimizing application performance.

---

[53] https://doc.zih.tu-dresden.de/hpc-wiki/bin/view/Compendium/SystemTaurus
[54] http://www.cscs.ch/computers/index.html
[55] https://sites.google.com/site/monitoringlargescalehpcsystems/
[56] https://www.hpc.uni-freiburg.de/nemo